# Extensions for Shared Resource Orchestration in Kubernetes to Support RT-Cloud Containers

Gabriele Monaco, Gautam Gala, Gerhard Fohler
Technische Universität Kaiserslautern, Germany
{monaco,gala,fohler}@eit.uni-kl.de

*Abstract*— Industries are considering the adoption of cloud computing for real-time applications due to current improvements in network latencies and the advent of Fog and Edge computing. To create an RT-cloud capable of hosting real-time applications, it is increasingly significant to improve the entire stack, including containerization of applications, and their deployment and orchestration across nodes. However, state-of-the-art orchestrators (e.g., Kubernetes) and underlying container engines are designed for general purpose applications. They ignore orchestration and management of shared resources (e.g. memory bandwidth, cache, shared-interconnect) making them unsuitable for use with an RT-cloud. Taking inspiration from existing resource management architectures for multicore nodes, such as ACTORS, and for distributed mixed-criticality systems, such as the DREAMS, we propose a series of extensions in the way shared resources are orchestrated by Kubernetes and managed by the underlying Linux layers. Our approach allows fine-grained monitoring and allocation of low-level shared resources on nodes to provide better isolation to real-time containers and supports dynamic orchestration and balancing of containers across the nodes based on the availability and demand of shared resources.

## I. INTRODUCTION

As seen in Ericson's Computing Fabric [1], and EU SE-CREDAS project [2], industries are exploring cloud computing for real-time applications to benefit from ease of re-usability, maintainability, and reconfiguration while providing workload elasticity and higher availability. However, the adoption of cloud computing for real-time applications is hampered by performance uncertainties as the current cloud models are not designed to guarantee worst case requirements. Some of the major hurdles are achieving high resource utilization in the cloud while providing timing guarantees and low interference between users, and supporting dynamic workloads and fluctuations in infrastructure availability. Previous work, such as [2], [3], have proposed using private RT-clouds with virtualization solutions and resource management extensions to host real-time and best-effort virtual machines in the cloud. However, virtualization solutions tend to have high overheads. In contrast, the overhead of containers compared to direct execution on top of the operating system is extremely low, making them a lucrative option for real-time applications.

Containers are becoming the de facto industrial approach, especially in cloud environments as they provide an ideal underlying layer for edge-to-cloud and multi-cloud scenarios. Organizations leverage containers to use multiple public cloud providers as well as their on-premises cloud. Container orchestrators, such as Kubernetes (K8s) [4], manage, deploy and scale containers across clusters of physical and virtual nodes. Orchestrators allow easy deployment of containers by properly setting up their runtime and monitoring their execution without the requirement of a cumbersome and error-prone manual process. In cloud environments, containers are often deployed via such orchestrators, managing certain resources, such as CPU-time and memory space, and delivering acceptable Quality-of-Service (QoS) to different users, typically based on their subscription plans. Based on the user configuration, orchestrators deploy containers to balance the usage of certain resources across the cluster nodes and guarantee these resources to the containers. Moreover, orchestrators can be configured to have container replicas, which in case of node failures, can be deployed on different nodes.

Orchestration of cloud resources is gaining relevance for the real-time systems, especially due to the advent of Fog and Edge computing, which brings the nodes closer to the end-users, thus, considerably lowering the latencies and enabling time-sensitive containers in the cloud. Orchestrators can help configure containers to exploit resource isolation possibilities provided by the underlying hardware and software layers to support meeting real-time requirements. However, widely used resource orchestrators (e.g., K8s) and the underlying Linux container technology (e.g. Docker [5]) and resource management approaches (e.g. Cgroups [6]) are not designed to consider strong shared resource isolation and end-to-end guarantees. They typically aim to improve average-case performance without regard for the worst case. Thus, the level of isolation provided by containerization and supported by container orchestrators is not adequate for real-time systems. Although optimizations and patches to reduce latencies are available, it is challenging to meet end-to-end constraints by considering only temporal allocation of CPU and spatial allocation of memory. A system with dynamically changing availability and requirement of resources requires the orchestrators to be aware of other resources of the node (e.g., memory bandwidth, cache, and shared interconnect) and services running in the container. The orchestrators must coordinate system-wide resources and adapt the QoS of services in the containers to the current resource availability (and not suspend containers, even the best effort ones, as far as possible).

Existing cloud orchestrators (e.g. K8s) work well when considering temporal allocation of CPU and spatial allocation of memory space, but do not support other shared resources. However, they require several extensions to ensure strong shared resource isolation and end-to-end guarantees under varying operating conditions. Existing works, such as [7], enable some real-time capabilities in Kubernetes and underlying technologies. Cinque et al. [8] enable containers and the Linux kernel to react with real-time performance. Such works aim to prevent over-reservation of CPU time to containers, indirectly targeting the weak isolation intrinsic to containers. Other works, such as [9], [10], exploit the modularity of Kubernetes to extend its capabilities over more resource types (e.g. network bandwidth), allowing better resource allocations decisions, but still do not consider allocation and monitoring of shared resources. Contrarily, our solution considers the allocation of shared resources and relies on low overhead run-time monitoring of hardware and software at various levels to obtain a system-wide view of availability of resources and current/predicted demand of containers, allowing the orchestrator to meet the end-to-end requirements of real-time containers and help achieve the best possible QoS by best-effort containers.

In this paper, we consider Kubernetes as the currently most popular container orchestration technology [4], We take inspiration from existing resource management architecture for multicore systems, such as ACTORS framework [11], and for distributed mixed-criticality systems, such as the DREAMS [12], to extend it. The existing ACTORS framework only considers a single multicore node, while DREAMS framework considers offline allocation of shared resources to virtual machine in a distributed system. However, offline allocation of resources is unrealistic in modern clouds. In fact, clouds require orchestrators to dynamically provide resource and maximise their utilization, and add new containers or provide scalability to existing ones at run-time. We build upon the DREAMS node-level (local) and system-wide (global) resource managers to extend the dynamic resource orchestration capabilities of Kubernetes while still ensuring strong resource isolation and fault tolerance. We aim to add configurable monitoring capabilities to Kubernetes at various levels ranging from low-level hardware signals (e.g., performance monitor counter events), operating system level events, and events in the containers, so that the orchestrator can keep a system-wide view of availability of shared resources and current/predicted demands of services running in the containers. Based on this system-wide view, the orchestrator can allocate resources to new containers or re-allocate resources to existing containers. The aim is to support real-time applications in clouds and ensure that their resource demands are promptly met while best-effort containers achieve the best possible QoS.

Section II introduces the DREAMS resource management and the standard resource orchestration in Kubernetes. Section III illustrates previous works that extended Kubernetes and Linux containers to support real-time containers. Section IV provides a series of enhancements for the orchestrator and the underlying layers to orchestrate and manage real-time containers together with best-effort containers. We describe how these enhancements can be applied in Kubernetes (with Linux) and finally conclude in Section V, providing future research and implementation directions.

## II. BACKGROUND

### A. DREAMS resource management architecture

We identified the resource management architecture delineated by the DREAMS project as a good base for creating extensions to the Kubernetes container orchestrator and the underlying Linux layers for supporting RT-Cloud. Several works are based on and extend this architecture [13], [3]; hence we can take benefit of those as well.

The resource management architecture delivers fault-tolerance and ensures real-time guarantees in mixed criticality distributed systems, where each node is capable of running multiple real-time and best-effort VMs. It attempts to provide best-effort applications with the best QoS level without compromising real-time guarantees in the system.

The DREAMS architecture decouples the system-wide orchestration decision making from the fine-grained resource management on the nodes. Each node runs a Local Resource Manager (LRM), capable of monitoring and scheduling the resources of the node. Furthermore, a Global Resource Manager (GRM) controls and coordinates the LRMs. The GRM takes decisions related to the entire system. By knowing the abstract sysem status on each node via the LRMs, the GRM can react to failures by reconfiguring nodes, and improves the resource balancing when overloaded situation occurs in a node. The capabilities of LRMs are split among the two subcomponents: monitors (MON) and Local Resource Schedulers (LRSs). The MON is responsible for observing the utilization of resources, and status and health of applications and resources. Examples of MONs are core failure monitor or performance monitoring event for CPU cache misses. The LRS schedules and controls resources accesses from applications on the nodes. Examples of LRS are CPU scheduler, network bandwidth regulators or techniques such as cache partitioning.

### B. Kubernetes

Kubernetes is highly configurable. We can extend or even replace components without patching the original codebase. In this section, we will introduce the default resource management.

During configuration, each container can specify resource requests and limits. Requests are used for scheduling: containers will not be deployed to nodes that cannot guarantee the resource, based on declared requests of other containers. Limits are usually enforced with Linux cgroups, to avoid overbooking (use of more peak resources than requested). This optimistic approach relies on resources being used less than their requests, allowing containers to use more than their requests and effectively allocating more resources than available. A more predictable scenario uses limits to avoid this situation (overbooking) from negatively impacting critical tasks. Natively available resources are CPU and memory space.

Providers can define the so-called *extended resources*, designed for resources like external devices (GPUs, FPGAs, etc.), whose number is discrete and which cannot easily be shared. Specific device plugins can bridge the actual driver with Kubernetes, initializing devices and providing the orchestrator with useful information about the status and availability of the nodes. Containers requesting a device will be granted full access to it until all available devices have been requested. A device is hence consumed if a container requests it in its specifications. Due to this architecture, the system does not enforce limits: the requested amount of resources is all the containers can get, regardless of their actual usage.

## III. RELATED WORKS

Works in the literature employ Kubernetes with performance-constrained IoT devices as [14], where the authors integrated a light version of the orchestrator (K3s) with the FogBus2 framework. Struhár et al. [7] enriched Kubernetes' support for real-time hierarchical scheduling on nodes via Linux kernel patches. They tackle the poor isolation of containers by monitoring the execution of tasks (e.g., deadline misses and lateness) and adjusting with future scheduling decisions. Similarly, Fiori et al. [15] modified Kubernetes for awareness of real-time scheduling parameters on deployment. Parameters such as runtime on rt-cpus per period are checked during admission control and the task scheduler is configured accordingly. Xu et al. [9] implemented the network bandwidth as a resource in Kubernetes. The authors react on pod deployment and control the Linux kernel stack according to specific rules. By creating some differentiated queues on virtual interfaces, they can give guarantees and limit overbooked resources, still keeping the Kubernetes handling unmodified. Yeh et al. [10] extend the GPUs handling in Kubernetes. They create an additional controller that schedules the GPUs based on user requirements. This can set time and memory (just like CPU), the parallel scheduler ensures minimal fragmentation while sharing devices across containers. Struhár et al. [16] assess whether and how Linux containers can deliver real-time performances and identify gaps regarding their deployment in safety-critical scenarios. They identify three categories of ways to achieve this: `PREEMPT_RT` for Linux Kernel to improve its responsiveness, exploiting a real-time co-kernel running alongside Linux and scheduling with improved guarantees the real-time containers, and hierarchical scheduling of containers and their internal tasks. Cinque et al. [8] propose an implementation scheme for real-time-enabled containers running on a Linux kernel with the co-kernel patch RTAI. Their implementation enforces temporal separation among containers by using a fixed-priority scheduler, run-time execution monitoring and mitigation. In case of misbehaviour, they provide policies to prevent interference from lower priority containers. In [17], they improve the previous work with a dynamic EDF scheduler for more flexibility. Abeni et al. [18] based their work on hierarchical scheduling, patching the kernel to allow 2 scheduling levels, for both LXC containers and their tasks. The authors instantiate their work and implement supporting facilities in the OpenStack orchestrator [19], configuring their hierarchical scheduler from a modified version of the orchestrator to implement Network Function Virtualization services.

## IV. EXTENDING KUBERNETES AND LINUX CONTAINERS TO SUPPORT REAL-TIME APPLICATIONS

Kubernetes is designed to be extensible, it provides a default configuration and implementation of most of the required components, but it isn't limited to it. System designers can reimplement the pod scheduler to substitute or even assist the default scheduler's implementation. This allows extensions without the need to restructure nor patch the orchestrator. Due to its modularity, it relies on certain components, the Linux containers and the related Control Groups, to name a few. Such systems may not be ready to target real-time applications, but their improvement is transparent to Kubernetes.

Orchestration systems work on networks of distributed nodes, deployment latencies are hence typically high by design. However, for real-time tasks to properly work, the run-time (after deployment) must be interfered with as little as possible. Kubernetes relies on the underlining OS for most of the runtime checks and operations (e.g. task scheduling or resource control). We can therefore achieve real-time guarantees by both having each node real-time capable (i.e. running a real-time capable OS) and making the orchestrator aware of the relevant resources to consider during deployment.

In the following subsections, we will discuss some points where the orchestrator and its stack can be improved or extended, giving some possible implementation and design guidelines on how to do it.

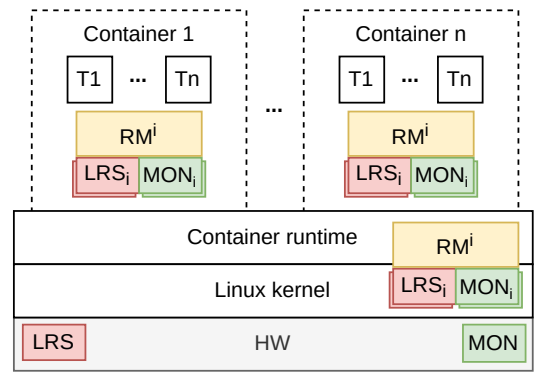

Fig. 1: Local resource managers in a node

### A. Resource monitoring

Most virtualization and orchestration solutions used in the cloud already support some resource monitoring and regulation,

accounting for runtime usage and preventing tasks to use more than reserved. Nevertheless, the considered items are usually only CPU time and memory space. Due to the shared nature of resources such as CPU cache, memory and bus, multicore architectures typically suffer from interferences. Those shared resources are generally ignored by the tools currently used in the cloud.

Works such as Memguard [20] propose a way to regulate the memory accesses by relying on data provided by the hardware Performance Monitoring Units (PMUs). Such units are widely supported on modern processor architectures and typically allow several low-level events to be counted. Raw monitoring data can be retrieved natively or with tools such as Linux perf. We hence employ the hardware PMU to monitor multiple resources using the same principle, exploiting the architecture support, for instance, memory bandwidth usage, shared bus and processor cache. Raw values can be obtained with low overhead, as produced by hardware components, however, the values alone are not always meaningful, counted events can greatly vary, making the resulting response inaccurate Moreover, frequent reactions to changes introduce high overheads, negatively affecting the system. Consequently, we abstract the resource values in significant bands, previously defined per resource. As discrete numbers, bands are easier to handle and we can smooth out changes between them with filters. By properly stating rules for transitions among resource bands, we can consistently reduce the overhead due to reacting to state changes, still without losing generality or precision.

Monitoring can be exploited to implement various features in Kubernetes and its orchestrator stack. Besides resource usage, monitoring instances can detect faults on components as well as on tasks and this knowledge can improve scheduling and orchestration decisions. Our monitoring solution is integrated into the Linux kernel and monitors separately each task. By following the scheduling events in the kernel, we can react to changes in the executing activity and we store values for each monitored process. Thereby, we have a separate view of each running task and we can enforce limits on a process basis.

*B. Shared resource isolation for containers*

Kubernetes relies on the container runtime to execute pods, however containers have weak isolation with respect to shared resources (memory, network or disk). As a lightweight alternative to Virtual Machines (VM), containers share the underlying kernel and OS. The simpler structure makes them even more vulnerable to interferences which can negatively and unpredictably affect the performance of critical tasks.

Essentially containers are traditional Linux tasks with a set of common properties to group them and the use of namespaces to emulate isolation. We can monitor the execution of such tasks and assign resource budgets to such groups. The typical use case of pods/containers is to host what is logically a single application, It can include multiple processes and even multiple binaries, it is however conceptually a *service*. We model containers as tasks and the resource monitoring targets them rather than processes. By exploiting our monitoring facilities, we derive the resource usages of processes in containers, we then throttle the processes whose container is running out of predefined budgets and we communicate to the GRM to take proper action. The architecture for resource managers on nodes and containers is depicted in Figure 1. Shared resource management is complementary to hierarchical scheduling used in previous works, as we cannot achieve strong timing guarantees without considering the impact of interferences on such resources.

*C. Integrating awareness of shared resources in Kubernetes*

The resources explicitly targeted by Kubernetes are memory and CPU, with possible extension to external devices. Our monitoring instances can provide an abstraction to various low level shared resources and with the use of filters can carefully distinguish between transient and continuous overload conditions. We designed the resource as divided into allocation levels, corresponding to the previously described resource bands. Pods can request a certain amount of levels, based on their need and the system will reserve them from the pool of available ones.

We implement this type of management natively in Kubernetes using extended resources, however, the system does not allow different values for requests and limits for such resources. Nevertheless, we exploit annotations in the containers specifications to indicate whether the requirements are strict (for real-time priorities) or loose (best effort), for instance, which we handle in our module. If global scheduling decisions take into account metrics such as available memory bandwidth, the main overhead of the orchestrator extension is during deployment (startup), while the control run-time logic is fully relying on the monitoring instances on the nodes. Both real-time and best-effort containers must declare their resource requirements and/or limits, also regarding, for instance, memory bandwidth. It is hence a responsibility of the system designer to both know the total available memory bandwidth and how can it be shared among the tasks of the system. We provide different policies such as throttling containers on the node or rescheduling them to a different node, based on the status and availability of the cluster. The architecture for resource managers across the cluster is depicted in Figure 2, our additional resource manager is shown in solid background.

*D. System-wide shared resource balancing*

Kubernetes allocates containers on nodes taking into account requested resources and nominal availability on nodes, once this check is passed, containers are assigned without further balancing or prioritization. Nodes can however receive a *score* after the feasible ones have been filtered. This value will be used to select the most suitable nodes for deployment, the scoring system is however not defined and we need to design it with a plugin to the scheduler.

Kubernetes doesn't implement seamless rescheduling, simple pods are either allocated on a node or deleted from it, with no possibility to move them across nodes. The closest functionality is to instantiate pods in controller structures (i.e. *deployments*), taking care of recreating the pod if terminated. Nonetheless, recreating evicted pods might imply losing the entire job executed by the pod, unless some architectural refactoring prevents this condition. Pods can be automatically evicted as resource consumption on nodes crosses certain thresholds, such as out of memory or storage conditions. Yet, this can only prevent OOM conditions and not to improve the cluster performances by balancing resource-intensive pods. Resources such as memory bandwidth are more likely to run out compared to storage, and their effect is usually more relevant for the performances of RT tasks. It is hence crucial to be able to keep a correct balance among nodes before overload conditions happen.

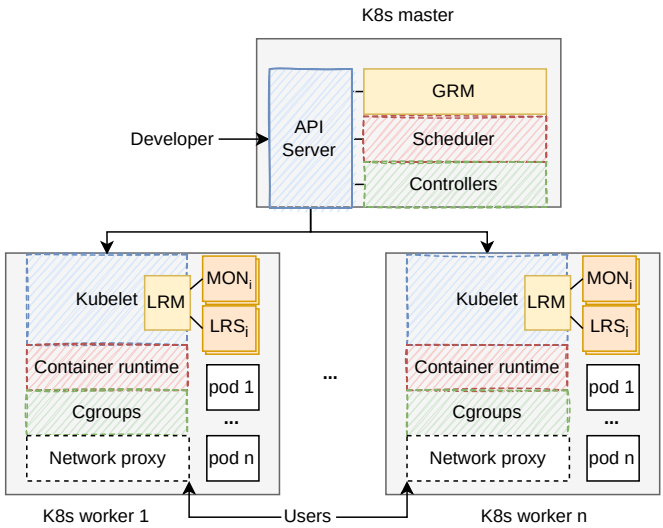

Fig. 2: Global and local resource managers with Kubernetes

We design different policies exploiting Kubernetes' scoring system, for instance, gathering best effort containers on the same nodes or prefer allocation on empty nodes for resource-intensive tasks and allocating tasks with lower resource requirements on busy nodes. We rely on live monitoring for further runtime checks and reorchestrate if the status is different from the predicted one or requested during deployment. Live monitoring can detect poorly configured containers and require readjustments before they unbalance the system. As rescheduling to a different node can result in high latencies due to network communication, we need to select with care the containers undergoing the process. Those can be less critical tasks or, collaboratively, tasks advertising their availability to migrate.

*E. System overload monitoring*

Kubernetes is primarily designed for resource usage maximisation rather than for guarantees to critical tasks. While this works during general usage, during overload, priority classes don't provide deterministic guarantees for real-time tasks, as limits enforcement and regulation can be slow. Kubernetes allows to either have all containers exposing same values for limits and requests, preventing overload by design, or better exploit the resources on the system but suffer from unpredictability. The first case is predictable as we set our requests to the worst case requirement. This can result in poor resource usage, as most applications don't consume their worst case requirements all the time. The less conservative approach allows the limit value to mirror the worst case usage, with requests corresponding only to best or average cases. Assuming the spikes in resource consumption on nodes don't occur at the same time, the system can use its resources more efficiently.

Monitoring the status of resources during runtime we can identify overload situation in advance and enforce stricter regulations only then. We allow hence resource overload without penalties in the general case and dynamically restrict it only when it can affect critical tasks.

*F. Collaborative QoS reduction on overload*

The default policy of Kubernetes in overload situations is to throttle down tasks (in case of CPU) or kill them (if running out of memory). The likelihood of these measures depends on the pod's priority class, defined by the pattern of request/limits (e.g. if both values are equal, the priority is higher). Applications are unaware of this and cannot adapt preventively (e.g. by internally throttling their QoS) nor be reallocated

To avoid uncontrolled degradation, we propose an interface for tasks to programmatically lower their resources footprint and readapt before the overload even occurs. We can implement this with scheduler extensions, starting a communication to tasks that request this functionality before corrective actions are taken, then if the overload condition is still present, the standard scheduler can take action.

*G. Dynamic resource orchestration*

Kubernetes provides resource requesting and limiting on pod creation, this cannot be changed dynamically as it is mainly relevant during allocation. Allowing limits higher than requests lowers the priority class. If critical applications change their requirement during their execution, they need either to have their requests mirroring the worst case or to join the Burstable class and have their resources guaranteed only in the best case.

We design additional interfaces to allow applications to change their resource requirements at run-time, triggering all required rescheduling on the orchestrator's side. A secondary pod scheduler in Kubernetes can be used to implement such interfaces.

*H. Node priorities and orchestration in heterogeneous clusters*

The default Kubernetes scheduler assumes all nodes are equivalent and schedules new workloads on the first available node according to its requirements. As mentioned earlier, user defined scoring systems can leverage that by sorting preferred nodes in case multiple options are available. For example, static prioritisation can help in a cluster with nodes powered by different energy sources, the owners might prefer to run and reschedule on nodes powered by green energy whenever possible, using the others only as a secondary option. In a cluster with heterogeneous nodes, with some driven by power conservative but weaker processors and others with stronger computing capability, a scoring system can exploit the CPU requirements of each container and deploy them accordingly.

We can enhance this pattern with live monitoring: gathering data from containers to understand which of them are either power intensive or need more CPU and reschedule them dynamically to optimize distribution in the aforementioned scenarios.

*I. Control for scheduling classes on nodes*

Kubernetes doesn't control the actual scheduling of tasks, just their assignment to nodes, then the OS scheduler will do the job. Some workloads, especially safety critical ones could take benefit from having a different kind of scheduling properties on nodes, such as a time triggered with slot shifting. The admission control of those pods can also compute a new scheduling table for the supported nodes and send it with the new task, ready to be scheduled in the end of the hyperperiod and with reduced overhead on the worker node. The nodes classes can be seen as a meta-resource by Kubernetes, which would activate a secondary OS scheduler for this task.

## V. CONCLUSION

Shared resources are ignored by popular container engines and orchestrators, making them unsuitable for use with a real-time(RT)-cloud. In this paper, we proposed extensions to Kubernetes and the underlying container engine for shared resource orchestration and management to support containers of RT-cloud. Our extensions are inspired by the DREAMS and ACTORS resource management architectures. We provided configurable monitoring capabilities at various levels so that the orchestrator can keep a system-wide view of the availability of shared resources and current/predicted demands of services running in the containers. Based on this system-wide view, Kubernetes can now dynamically orchestrate resources to execute new containers or dynamically balance containers across the nodes based on the current availability and demand of the shared resources. Moreover, the extensions to the underlying Linux layers ensure strong isolation in shared resources. The overall goal of our extensions is to ensure that real-time applications meet their resource demands while best-effort applications achieve the best possible QoS.

We follow the philosophy of Kubernetes of keeping all components modular and extensible. Our proposed design is transparent to different algorithms and strategies. Based on the requirements, system designers can select the monitoring and scheduling components that best fit their needs and plug them into our resource managers. Thus, we reduce the integration complexity without sacrificing flexibility and exploiting the full potential of Kubernetes.

Future steps involve completing the implementation of the proposed extensions and deploying this extended Kubernetes-based RT-cloud with industrial use cases to assess and evaluate the improvements over vanilla Kubernetes. Then by following the design principle in DREAMS during implementation, we can effectively guarantee end-to-end requirements of critical containers.